\documentstyle[12pt,a4,amssymb,epsfig]{article}
\def\d{\delta}
\begin{document}

\begin{flushright}
MPI-PhT/97-53\\
September 1997\\
\end{flushright}
\begin{center}
\large {\bf New Interactions in Top Quark Production and Decay at the 
Tevatron Upgrade} 
\\
\mbox{ }\\
\normalsize
\vskip1cm
{\bf Bodo Lampe}
\vskip0.2cm
Max Planck Institute for Physics \\
F\"ohringer Ring 6, D--80805 Munich \\
\vspace{1cm}

{\bf Abstract}\\                          
\end{center}                              
\vspace{-0.5cm}
New interactions in 
top-quark production and decay are studied under the conditions
of the Tevatron upgrade. Studying the process 
$q\bar q \rightarrow t\bar t\rightarrow b\mu^+ \nu\bar t$,  
it is shown how the lepton rapidity and transverse energy 
distribution are modified by nonstandard modifications of the 
$gt\bar t$-- and the $tbW$--vertex.  

\vspace{1cm}
\vspace{1cm}

Heavy particles like the top--quark provide an
interesting opportunity to study physics beyond the Standard Model 
because it is conceivable that 
nonstandard effects appear first in interactions of the 
known heavy particles 
(the top quark and the heavy gauge bosons). 

In this letter the process 
$q\bar q \rightarrow t\bar t\rightarrow bW^+\bar t\rightarrow b\mu^+ \nu\bar t$ 
will be studied assuming that it proceeds as in the Standard Model 
($t\bar t$ production by s--channel gluon exchange 
and subsequent decay to $bW$). 
We shall assume that all nonstandard effects in the 
production process $q\bar q \rightarrow t\bar t$ can be represented 
by modifying the $gt\bar t$ vertex. Similarly, nonstandard effects 
in the decay of top quarks will be parametrized by 
modifying the Standard Model $tbW$ vertex. 
Note that the $\bar t$ state is assumed to decay hadronically 
and its decay products are averaged over.  
Among all top quark events, these processes 
are particularly interesting because they show the best compromise 
between statistics and event signature. 
In fact, for a hadronic decay the $\bar t$ momentum can be 
fully reconstructed to fulfill $p_{\bar t} ^2=m_t^2$. This together 
with a hard lepton used as a trigger is a rather unique 
signature of top quarks in proton collisions. 
Furthermore, 
a refined analysis of production and decay dynamics is possible, 
because the $b$, the $l^+$ and the $\bar t$ momentum 
can be experimentally determined.

The most general effective $gt\bar t$ vertex can be parametrized 
as follows 
\begin{equation}
{\Gamma}^{\mu a} (g^{\ast} \rightarrow t\bar t)=
ig_{s}  \bar{u}(p_t) 
\biggl[\, 
\gamma^\mu (1+\d A_P-\d B_P\gamma_5)+
\frac{p_t^\mu-p_{\bar t}^\mu}{2m_t} 
(\d C_P-\d D_P\gamma_5)  \,\biggr] {\lambda^a \over 2}v(p_{\bar t} ) 
\label{g11}                               
\end{equation}    
where $g_{s}$ is the strong coupling constant and $\lambda^a$ 
the Gell--Man $\lambda$--matrices. 
The SM vertex is given by $\d A_P=\d B_P=\d C_P=\d D_P=0$. 
Note that there is an equivalent parametrization of the 
vertex by 
\begin{equation}
{\Gamma}^{\mu a} (g^{\ast} \rightarrow t\bar t)=ig_{s}
  \bar{u}(p_t)
\biggl[\,\gamma^\mu (F_1^L P_L+F_1^R P_R)
-\frac{i \sigma^{\mu \nu}(p_t+p_{\bar{t}})_\nu}{m_t}(F_2^L P_L+F_2^R
P_R)\,\biggr] {\lambda^a \over 2}v(p_{\bar t} )
\label{gtt}
\end{equation}
where $P_{L/R}=(1\mp\gamma_5)/2$.
Using the Gordon decomposition one can indeed show that 
$\d A_P={1\over 2}(F_1^L+F_1^R)-1-F_2^L-F_2^R$, 
$\d B_P={1\over 2}(F_1^L-F_1^R)$, 
$\d C_P=F_2^L+F_2^R$ and $\d D_P=F_2^L-F_2^R$.

Similarly, 
the following parameterization of the $tbW$ vertex
suitable for the decay $t\rightarrow bW^+$ will be adopted 
\begin{equation}
{\Gamma}^{\mu}(t\rightarrow bW^+ )=-i{g\over\sqrt{2}}V_{tb}  
\bar{u}(p_b)\biggl[\,
\gamma^{\mu}(P_L+{\d A_D \over 2}-{\d B_D\over 2}\gamma_5) 
+\frac{p_b^{\mu}+p_t^{\mu}}{2m_t} 
(\d C_D-\d D_D \gamma_5)   \,\biggr]u(p_t)
\label{w10}
\end{equation}
where $g$ is the SU(2) gauge-coupling constant and $V_{tb}$ the $(tb)$
element of the CKM matrix. 
The SM vertex is given by $\d A_D=\d B_D=\d C_D=\d D_D=0$.
A Gordon decomposition similar to the above leads to 
an equivalent description 
\begin{equation}
{\Gamma}^{\mu}(t\rightarrow bW^+ )=-i{g\over\sqrt{2}}V_{tb}
\bar{u}(p_b)\biggl[\,\gamma^{\mu}(G_1^L P_L +G_1^R P_R)
-{{i\sigma^{\mu\nu}(p_t-p_b)_{\nu}}\over m_t}
(G_2^L P_L +G_2^R P_R)\,\biggr]u(p_t)
\label{w11}
\end{equation}
Indeed, one has $\d A_D=G_1^L+G_1^R-1+G_2^L+G_2^R$, 
$\d B_D=G_1^L-G_1^R-1-G_2^L+G_2^R$,
$\d C_D=-G_2^L-G_2^R$ and $\d D_D=-G_2^L+G_2^R$.
Note the factor $m_t$ appearing in Eq. (\ref{w11}) 
whereas in Refs. \cite{grz,kane} the W--mass was used to 
normalize the nonstandard couplings $G_1^L$ and $G_1^R$. 

In Eqs. (\ref{g11})--(\ref{w11}) all terms have been neglected,  
which in the cross section 
give contributions proportional to the light fermion 
masses or to the off-shellness of the $W$--boson.  
Apart from such terms, Eqs. (\ref{g11}) and (\ref{w10}) 
comprise the most general interactions of top quarks with 
gluons and W--bosons, respectively. 

Using Eqs. (\ref{g11}) and (\ref{w10}), the matrix elements $M_P$ 
for $t\bar t$ production as well as $M_D$ 
for the decay process $t\rightarrow bl^+\nu$ and for the combined 
production and decay process 
$q\bar q \rightarrow t\bar t\rightarrow \bar tbl^+\nu$ 
were calculated. Only contributions linear in the 
nonstandard couplings were kept. One finds 
\begin{eqnarray} \nonumber
M_P&=& 
        [s^2+2m_t^2s-4s(p_t\cdot p_{q})+8(p_t\cdot p_{q})^2]
                             (1+2\d A_P)  \\  
 & &   +4\d C_P[m_t^2s-2s(p_t\cdot p_{ q})+4(p_t\cdot p_{ q})^2]
\label{mep}
\end{eqnarray}
where $s=(p_q+p_{\bar q})^2$ is the total partonic energy 
and $p_t\cdot p_{q}$ is  
related to the top quark production angle $\theta$ in the 
parton--cms : $2p_t\cdot p_{q}={s\over 2}(1-\sqrt{1-{4m_t^2\over s}}\cos 
\theta )$. The notation of Ref. \cite{combridge} was used and 
a factor ${64\pi^2\alpha_s^2\over 9s^2}$ has been left out.  
As is obvious from this equation, the coupling $\d A_P$  
renormalizes the total cross section whereas $\d C_P$ modifies 
the angular distribution and $\d B_P$ and $\d D_P$ do not 
contribute at all. Note, however, that $\d B_P$ and $\d D_P$ 
will strongly contribute -- via spin terms -- to the combined process 
$q\bar q \rightarrow t\bar t\rightarrow \bar tbl^+\nu$ (see below)! 
  
For the decay matrix element one finds 
\begin{eqnarray} \nonumber         
M_D&=&(p_t\cdot p_l)[{m_t^2\over 2}-(p_t\cdot p_l)](1+\d A_D+\d B_D) \\
& &  +(\d C_D-\d D_D)[-(p_t\cdot p_l)^2+{1\over 2}(p_t\cdot p_l)(m_t^2+m_W^2)
                   -{1\over 4}m_t^2m_W^2]
\label{med}
\end{eqnarray}
where $p_t\cdot p_l$ can be related to the lepton energy
$E_l$ in the rest system of the top quark : $p_t\cdot p_l=m_tE_l$. 
The notation and normalization 
of Ref. \cite{jezabek} was used. Obviously, the 
couplings $\d A_D$ and $\d B_D$ just renormalize the 
Standard Model cross section whereas $\d C_D-\d D_D$ 
really modifies the lepton distributions. 

The matrix element for the combined production and decay process 
is not just the product of Eqs. (\ref{mep}) and (\ref{med}), 
but contains additional terms $\sim \d B_P$ and $\sim \d D_P$, 
i.e. one has $M=M_PM_D+\Delta$, with 
\begin{eqnarray} \nonumber 
\Delta &=&4\d B_P (p_{\nu}\cdot p_b) 
 \biggl\{      [(p_l\cdot p_t) (p_q\cdot p_t)(p_{\bar q}\cdot p_{\bar t}) 
         -m_t^2(p_l\cdot p_{\bar q})(p_q\cdot p_{\bar t})]  
                +[q \leftrightarrow \bar q]  \biggr\} \\  
& &  +\d D_P (p_{\nu}\cdot p_b) \biggl\{ 
   [(p_q\cdot p_{\bar q})(p_l\cdot p_t)(p_t\cdot p_{\bar t})
     -m_t^2(p_q\cdot p_{\bar q})(p_l\cdot p_{\bar t})   
    +(p_l\cdot (p_t+p_{\bar t})) 
\nonumber \\  & & \times
       (p_q\cdot p_t) (p_{\bar q}\cdot (p_t-p_{\bar t})) 
     -(p_q\cdot p_{\bar q})(p_l\cdot p_{\bar q})(p_1\cdot (p_t-p_{\bar t}))  
  ] + [q \leftrightarrow \bar q]  \biggr\} 
\, .
\label{mec}
\end{eqnarray}
These latter terms arise when the 'spin contributions' $\sim s_t$ of the 
amplitude $A(q\bar q\rightarrow t\bar t)$ (i.e. the terms 
proportional to the spin vector $s_t$ of the top quark) 
are 'contracted' with 
the 'spin contributions' of the decay amplitude 
$A(t\rightarrow bl^+\nu)$ using $s_t^2=-1$. Note that such 
term are not present in the Standard Model. 
Spin terms arise in 
the Standard Model if correlations of t {\it and} 
$\bar t$ decay are considered \cite{parke}, or if there is 
an axialvector component of the Standard Model coupling 
on the production side, like in $e^+e^- 
\rightarrow t\bar t$ via Z--exchange \cite{arens,kane}. 
Note further that the terms $\sim \d D_P$ and $\sim \d D_D$ in the 
above expressions give rise to CP violating effects when 
the behavior of top and antitop quark is compared \cite{grz,cpv}. 

\begin{figure}
\begin{center}
\epsfig{file=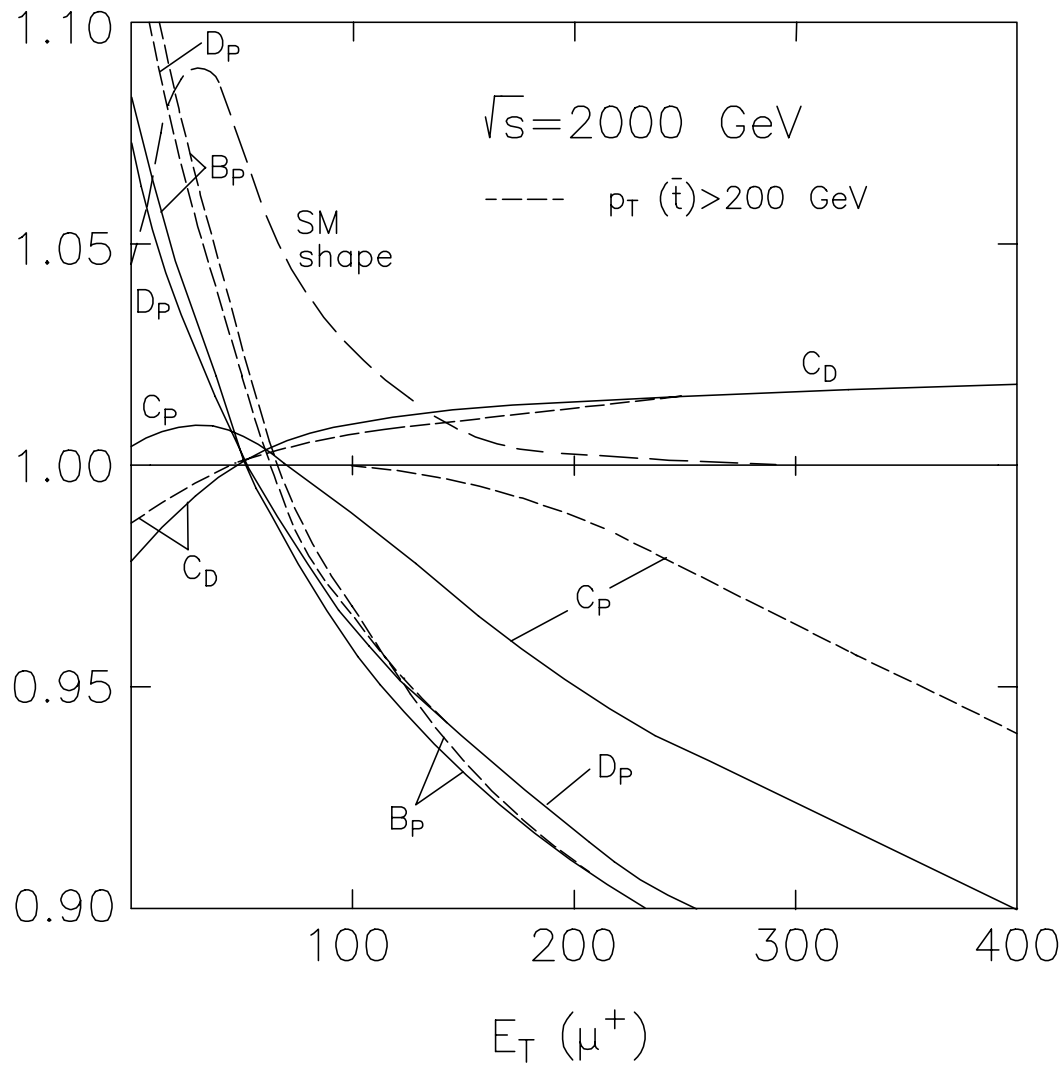,height=14cm}
\bigskip
\caption{   
The ratio of nonstandard to SM contribution as a function of 
the lepton ($l^+$) transverse energy, for various nonstandard 
terms denoted by $B_P$, $C_P$, $D_P$ and $C_D$, c.f. Eqs. 
(\ref{g11}) and (\ref{w10}). The values of the couplings 
were chosen to be 
$\d B_P=0.1$, $\d C_P=0.1$, $\d D_P=0.1$ and $\d C_D=0.1$. 
Also included is the shape of the SM contribution (in arbitrary 
units). 
}
\end{center}
\end{figure}

Using the matrix elements Eqs. (\ref{mep}), 
(\ref{med}) and (\ref{mec}) one can determine the lepton 
rapidity and transverse energy distribution under the 
conditions of the Tevatron upgrade. The Tevatron upgrade is 
defined by a total energy of $\sqrt{S}=2$ TeV and 
two options for the luminosity, the so called
`TeV-33' defined as $L=30$ fmbarn$^{-1}$ and the Tevatron Run II with
$L=2$ fmbarn$^{-1}$ \cite{upgrade}. 
The expected number of single-leptonic events (1 b-quark
tagged) \cite{upgrade} is
$1300$ and $20,000$ for $L=2,\; 30$ fmbarn$^{-1}$, respectively.
Numerical results were 
obtained using the Monte Carlo package RAMBO \cite{kleiss}. 
Standard CDF and D0 cuts were applied. 
The matrix element squared were 
convoluted with the Morfin and Tung \cite{morfin}
parton distributions (the 'leading order' set from the 'fit sl').
Finally the ratio of the results to the Standard Model predictions 
were taken. Figures 1 and 2 show these ratios for coupling values 
$\d B_P=0.1$, $\d C_P=0.1$, $\d D_P=0.1$ and $\d C_D=0.1$, respectively. 
Figure 1 shows the dependence on the lepton--$p_T$ 
and figure 2 on the lepton rapidity. 

As one would expect, nonstandard effects are roughly of the order 
of 5--10 \%. Effects are larger for the transverse energy 
than for the rapidity distribution. 
The most pronounced effects come from $\d B_P$ and $\d D_P$ 
at intermediate and high lepton $E_T$.  
From Figs. 1 and 2 it is apparent that 
the contribution $\sim \d C_D$ is relatively smaller than the other ones. 
This proves that effects from the decay vertex are harder to 
find than nonstandard effects at the production vertex. 
The figures also include the shape of the Standard Model predictions (in 
arbitrary units). The short--dashed curves in Fig. 1 are obtained if a 
$p_T$--cut on the 
$\bar t$ momentum is applied. 
Since the $\bar t$ momentum is experimentally known, 
the dependence on $p_T(\bar t)$ may be analyzed in order to 
separate the different nonstandard effects. For example, the 
contribution $\sim \d C_P$ depends strongly on $p_T(\bar t)$ whereas 
the others do not.    

\begin{figure}
\begin{center}
\epsfig{file=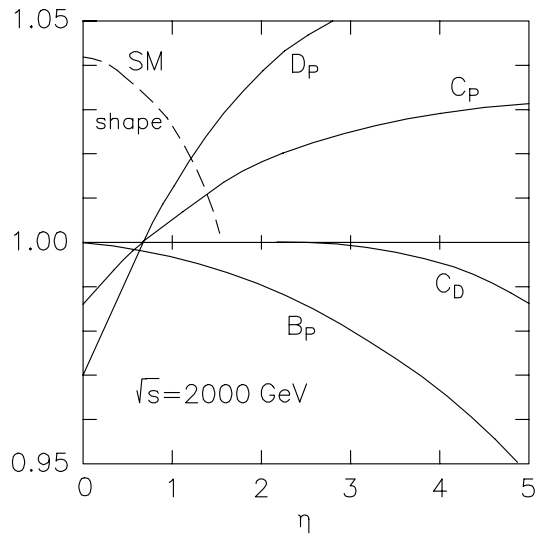,height=7cm}  
\bigskip
\caption{
Same as Fig. 1 but as a function of the $l^+$ rapidity.
}
\end{center}
\end{figure}

Using the results Eqs. (\ref{mep})--(\ref{mec}) it is also 
possible to calculate other distributions, like $p_T$-- and 
$\eta$--distributions for $\bar t$ and b--quark, or more 
complicated 2--particle correlations. As an example, 
Fig. 3 shows the ratio of nonstandard to SM contribution as 
a function of the angle $\phi$ between the transverse momenta of 
$\bar t$ and $l^+$. One sees, for example, 
that in the high--statistics region  
($\phi \sim \pi$) the interaction terms $\sim \delta B_P$ and 
$\sim \delta C_P$ can be clearly distinguished whereas the 
terms $\sim \delta B_P$ and $\sim \delta D_P$ give almost 
identical results. 

\begin{figure}
\begin{center}
\epsfig{file=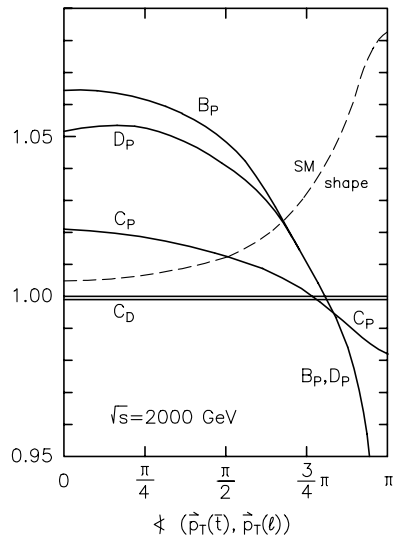,height=7cm}
\bigskip
\caption{
Same as Fig. 1 but as a function of the angle between the 
transverse momentum of $l^+$ and $\bar t$.
}
\end{center}
\end{figure}

In ref. \cite{jezabek1} the lepton energy 
distribution in top quark decays was analyzed including 
the nonstandard interactions Eq. (\ref{w11}). Since this 
was done in the rest system of the top quark, results are 
not directly comparable with the present analysis.  

To conclude, in 
this article I have calculated the full matrix elements as well 
as transverse energy and angular distributions 
for top quark production {\it and} decay under the 
conditions of the Tevatron upgrade. 
I have not included contributions from the process 
$gg \rightarrow t\bar t$ because they give less than 
10 \% of the top quark production cross section at Tevatron energies. 
Another approximation of the present letter is, that higher order 
QCD contributions have not been taken into account. These 
are in principle known because they are known for production 
and decay separately and spin terms do not 
contribute here. 
These contributions are also expected to be roughly  
of the order of 10 \% and are also needed for a 
precision analysis of future Tevatron data. 
I did not include them here because I just wanted 
to elucidate the role of nonstandard interactions 
with reference to the leading order standard model process. 

A more general aim of this paper is 
to point out, that nonstandard effects 
in top quark interactions may be found already before precision
measurements at the LHC will be done.

{\bf Acknowledgement. } \\                          
I am indebted to B. Grzadkowski for discussions on nonstandard 
top quark interactions.

\end{document}